\documentclass[aps,pra,twocolumn, superscriptaddress]{revtex4-1}

\usepackage{amsmath,amssymb}
\usepackage{xcolor}
\usepackage{graphicx}
\usepackage{braket}
\usepackage[utf8]{inputenc}

\begin{document}

\newcommand{\h}[1]{\hat{#1}}
\newcommand{\sinc}{\mathrm{sinc}}
\newcommand{\todo}[1]{\textcolor{blue}{ToDo: {#1}}}

\newcommand{\upb}{Integrated Quantum Optics Group, Applied Physics, University of Paderborn, Warburger Stra\ss e 100, 33098 Paderborn, Germany}
\newcommand{\prag}{FNSPE, Czech Technical University in Prague, B\v{r}ehov\'a 7, 115 19, Praha 1, Czech Republic}

\title{Dual-path source engineering in integrated quantum optics}

\author{Regina Kruse$^1$, Linda Sansoni$^1$, Sebastian Brauner$^1$, Raimund Ricken$^1$, Craig S. Hamilton$^2$, Igor Jex$^2$, and Christine Silberhorn}
\affiliation{ \upb\\
						 $^2$ \prag}
						 

\begin{abstract}
Quantum optics in combination with integrated optical devices shows great promise for efficient manipulation of single photons. New physical concepts, however, can only be found when these fields truly merge and reciprocally enhance each other.
Here we work at the merging point and investigate the physical concept behind a two-coupled-waveguide system with an integrated parametric down-conversion process. We use the eigenmode description of the linear system and the resulting modification in momentum conservation to derive the state generation protocol for this type of device. With this new concept of state engineering, we are able to effectively implement a two-in-one waveguide source that produces the useful two-photon NOON state without extra overhead such as phase stabilization or narrow-band filtering. Experimentally, we benchmark our device by measuring a two-photon NOON state fidelity of $\mathcal{F}=(84.2\pm2.6)\,\%$ and observe the characteristic interferometric pattern directly given by the doubled phase dependence with a visibility of $\mathcal{V}_{\mathrm{NOON}}=(93.3\pm 3.7)\,\%$.
\end{abstract}

\maketitle

\section{Introduction}
Compact, low-cost, and easy-to-use devices are big advantages of the mature field of integrated optics. The large number of well-engineered devices and achieved re-configurability of the implemented circuits allow for fast and efficient manipulation of light. These highly desired properties have enabled the quantum information community to build and investigate compact, efficient, and high-dimensional networks for the first time \cite{poli08sci, cres11nco, corr14nco, heil13sre, peru10sci, sans12prl, broome_photonic_2013, crespi_integrated_2013, spring_boson_2013, tillmann_experimental_2013}. Still, the generation of single photons that fuel the optical circuits is usually realized off chip with bulk sources of parametric down-conversion (PDC) \cite{boyd_nonlinear_1992}. While these offer high flexibility in the spatial domain \cite{walborn_spatial_2010}, the low brightness compared to their waveguide counterparts \cite{fiorentino_spontaneous_2007}, as well as possible losses and instabilities in the bulk-waveguide interface, prevents the scaling to many photons coupled to many inputs of a linear network.
Waveguide sources generate the photon pairs into fiber-compatible spatial modes \cite{mitomi_design_1994}, allow for engineering of the spectral properties \cite{saleh_modal_2009, harder_optimized_2013}, and increase the conversion efficiency due to the high confinement of light \cite{herr13oex,tanzilli_highly_2001}. However, this big advantage is also the downfall of this approach, as the confinement prohibits the engineering of the path degree of freedom, which is easily accessible in bulk sources.
One class of states that makes exemplary use of the path degree of freedom is the NOON states \cite{boto00prl}. These maximally path-entangled states given via $\ket{NOON}=\frac{1}{\sqrt{2}}(\ket{N0}+e^{i\varphi}\ket{0N})$ exhibit an enhanced phase sensitivity that depends linearly on the photon number $N$. This property defines the main applications of NOON states as quantum metrology and lithography \cite{dowling_quantum_2008,boyd_quantum_2012}. However, the experimental implementation directly relies on the path degree of freedom \cite{afek_high-noon_2010, israel_experimental_2012}, making an integrated realization difficult.
Only last year, two devices were demonstrated \cite{silv14npo, jin14prl} that gain control over the needed path degree of freedom, using a workaround by combining a pump beam splitter with a phase shifter and two single waveguide sources. While this approach is successful in the end, it uses additional resources, such as postselection and narrow-band filtering. The natural follow-up question to their results is whether we can find a physical concept that \textit{intrinsically} addresses two waveguide outputs and therefore introduces the path degree of freedom in waveguide based technologies.
An interesting candidate to solve this problem is periodically poled waveguide arrays \cite{solntsev_spontaneous_2012, solntsev_generation_2014, wu_photon_2014}, which have been exploited for driven quantum walks \cite{hamilton_driven_2014}. In this context, it is known that the geometric layout of the coupled waveguides has an impact on the governing phase-matching function of the nonlinear PDC process \cite{kruse_spatio-spectral_2013}.
However, the implications of this new type of device are certainly not obvious, due to the large scale and many degrees of freedom. It is therefore necessary to investigate smaller systems with an integrated nonlinearity  \cite{mista_nonclassical_1997, lugani_generation_2011}, as less degrees of freedom enhance the chances for a clean and easy to manipulate system. While the properties of a single-waveguide source of PDC are well known and the calculation algorithms for large waveguide arrays are understood, the features of a coupled two-waveguide system with integrated PDC are still an open research question.

Here we explore the physics and generation concept of a nonlinear two-coupled waveguide geometry and show that, with this approach, we gain intrinsic control over the path degree of freedom with this approach. The combination of the eigenmodes of the linear coupled structure with the momentum conservation of the nonlinear process allows us to go beyond reproducing a bulk setup on a chip and to effectively implement a two-in-one waveguide source with a single optical element. In detail, we exploit the spatio-spectral coupling between the pump wavelength and selection rules for the eigenmodes and find that a Hong-Ou-Mandel effect in the transformation between the generation in the eigenmodes and the detection in the waveguide basis generates the useful two-photon NOON state. This elegant generation concept allows us to eliminate the need for phase-stable multiwaveguide pumping and narrow-band filtering. Furthermore, it is largely independent of fabrication parameters and imperfections.
Experimentally, we demonstrate the selectivity of the state-generation concept in the spatial domain by measuring the spatially resolved photon pair correlations depending on the pump wavelength and verify the phase coherence by observing the expected double-fringe pattern of a two-photon NOON state.

Our paper begins with a discussion about the theoretical framework of our device in Sec. II. In Sec. III we provide details on the chip design and the source parameters, before presenting the experimental setup and the results in Sec. IV. We summarize our findings in Sec. V.

\section{Theory}
\begin{figure}
\includegraphics[width=.9\columnwidth]{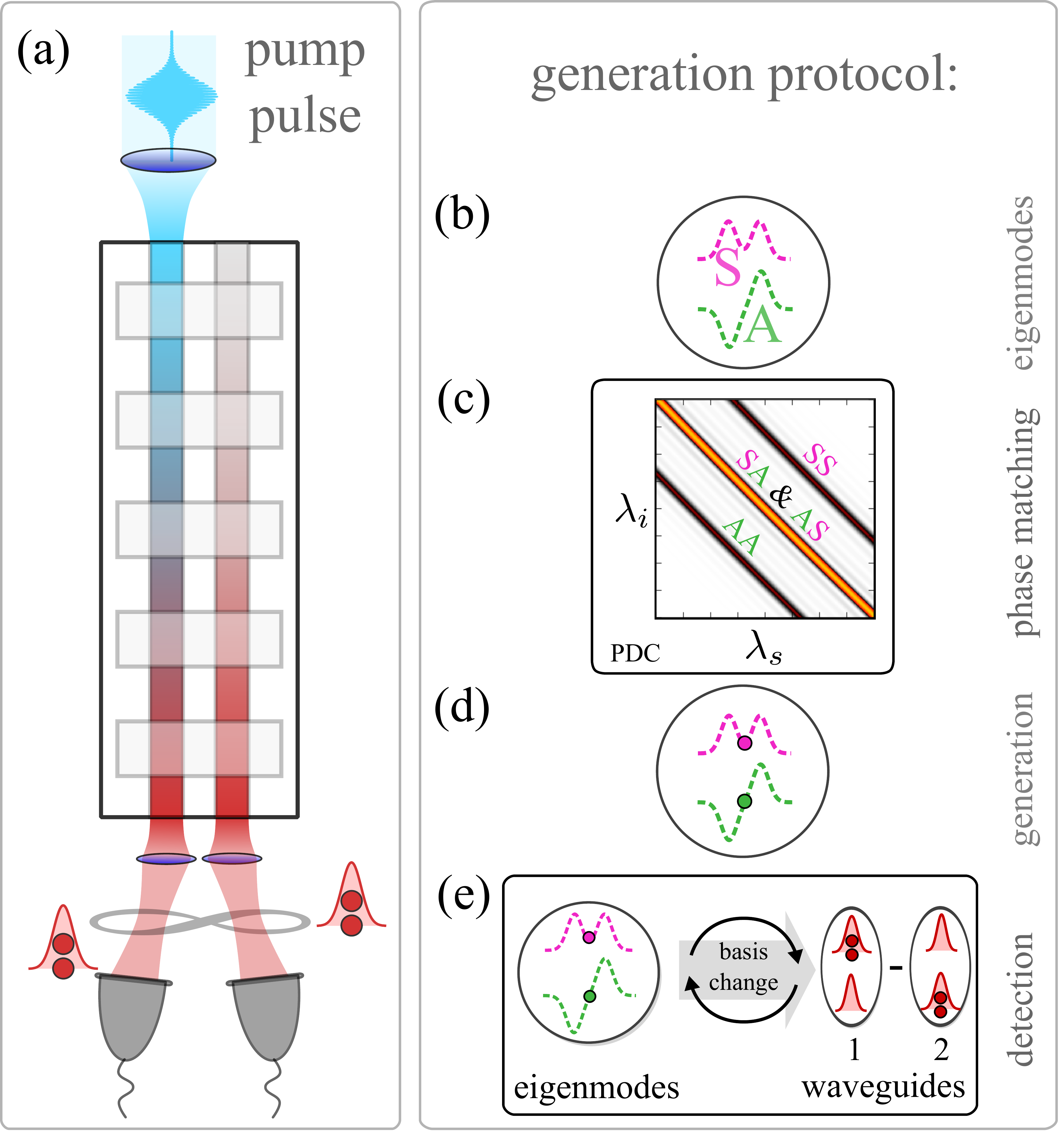}
\caption{(Color online) System and generation protocol. (a) The source consists of two periodically poled coupled waveguides and we can generate a two-photon NOON state by pumping a single waveguide. (b) The coupled structure has two eigenmodes (symmetric $S$ and antisymmetric $A$) with nondegenerate eigenvalues. (c) Different distributions of the PDC photons across the eigenmodes lead to a splitting of the phase-matching function. (d) By selecting the central phase matching, we generate a photon in each eigenmode. (e) Finally, the transformation from eigenmode to waveguide basis results in a two-photon NOON state. 
}
\label{fig:sys_sketch}
\end{figure}
We consider a system as sketched in Fig. \ref{fig:sys_sketch}(a). It consists of two waveguides that run parallel with a separation distance of a few micrometers (for detailed information concerning the sample parameters refer to Sec. III). The strength of the coupling, described by the coupling parameter $C$, is directly given by the distance between the two waveguides and the operating wavelength of the directional coupler. We design the waveguides such that only fields in the telecom regime are affected by the coupling geometry while near infrared light remains undisturbed. Additionally, we add a periodic poling of length $L$ to the coupling region, which enables the nonlinear PDC process \cite{louisell_quantum_1961, burnham_observation_1970} at a chosen wavelength combination.

The full Hamiltonian of the poled coupler system is given by a linear part given by the free propagation and the coupling behavior of the fields, as well as a nonlinear interaction part describing the PDC process
\begin{equation}
\h{H}_{\mathrm{PDC}} = \chi^{(2)}\int_V \mathrm{d}^3r\, (\mathcal{E}_p^{(+)} \h{E}^{(-)}\h{E}^{(-)} + H.c.) \, ,
\end{equation}
where $\chi^{(2)}$ is the effective nonlinear coefficient of the system. As we treat the pump field $\mathcal{E}_p$ as classical, only the generated fields are described by operators. We do not use subscripts for the quantum fields, as the photons are fundamentally indistinguishable and therefore described by the same operator. To solve the full system Hamiltonian including the coupling, we can express the fields in the interaction part in the eigenmode picture \cite{somekh_channel_1973, marom_relation_1984}. This way, the solution to the linear part is already included in the formulation of the nonlinear part. In the case of two coupled waveguides, there are two nondegenerate eigenvalues $\beta_{S(A)}=\beta^{(0)}\mp C$, where $\beta^{(0)}$ is the propagation constant of the uncoupled system. The labels $S$ and $A$ refer to the shape of the symmetric [sketched magenta in Fig. \ref{fig:sys_sketch}(b)] and antisymmetric (green) eigenmodes. The transformation between the waveguide and eigenmode picture is given by a linear combination of the uncoupled waveguide modes $\h{a}^\dagger_{S(A)}(\omega)=\frac{1}{\sqrt{2}}(\h{b}^\dagger_1(\omega)\pm \h{b}^\dagger_2(\omega))$, where $\h{a}^\dagger_k(\omega)$ creates a photon of frequency $\omega$ in the eigenmode $k$ and $\h{b}_j^\dagger(\omega)$ denotes the creation operator for a photon in waveguide $j$.

After a lengthy but straightforward calculation \cite{christ_spatial_2009, solntsev_spontaneous_2012, kruse_spatio-spectral_2013}, we arrive at the two-photon PDC state in the eigenmode basis of the waveguide coupler
\begin{equation}
\begin{aligned}
\ket{\Psi}^{\mathrm{eig}}&=\frac{1}{\sqrt{\mathcal{N}}} \int\mathrm{d}\omega_s \int\mathrm{d}\omega_i\, \alpha(\omega_s+\omega_i)\\
&\hspace{-0.1cm}\times\left[\right.\gamma\, \Phi(\Delta\beta_{SS},L)\, \h{a}_S^\dagger(\omega_s)\h{a}_S^\dagger(\omega_i)\\
&+\;\delta\, \Phi(\Delta\beta_{SA},L)\, \h{a}_S^\dagger(\omega_s)\h{a}_A^\dagger(\omega_i)\\
&+\;\delta\, \Phi(\Delta\beta_{AS},L)\, \h{a}_A^\dagger(\omega_s)\h{a}_S^\dagger(\omega_i)\\
&+\;\gamma\, \Phi(\Delta\beta_{AA},L)\, \h{a}_A^\dagger(\omega_s)\h{a}_A^\dagger(\omega_i)\left.\right] \ket{0}\,.
\end{aligned}
\label{eq:eig_state}
\end{equation}
The function $\alpha(\omega_s+\omega_i)$ represents the spectral properties of the pump and contains the energy-conservation condition $\omega_p=\omega_s+\omega_i$ (the subscripts refer to pump, signal and idler respectively). Momentum conservation is included in the phase-matching function $\Phi(\Delta\beta_{MN},L)=\sinc\left(\Delta\beta_{MN}\frac{L}{2}\right) e^{-i\Delta\beta_{MN}\frac{L}{2}}$, which depends on the length $L$ and the phase mismatch $\Delta\beta_{MN}=\beta(\omega_p)-\beta^{(M)}(\omega_s)-\beta^{(N)}(\omega_i)$, as given by the combination of different eigenmodes $(M,N)$. Here $\mathcal{N}$ is the normalization constant and $\gamma$ and $\delta$ are the excitation amplitudes of the symmetric and antisymmetric eigenmode, respectively. If we pump the two waveguides in the symmetric mode (the pump phase between two waveguides is 0) $\gamma = 1$ and for the antisymmetric configuration (the pump phase between two waveguides is $\pi$) $\delta= 1$. In the case of pumping only a single waveguide, both excitation amplitudes are equal. This multitude of accessible parameters leads to a high flexibility in the experiment.

The state in Eq. \eqref{eq:eig_state} contains four distinct phase-matching conditions, which we can identify as the four possibilities that we obtain by distributing two PDC photons across two eigenmodes. They possess spectrally separated phase-matching conditions, as the two corresponding eigenvalues are non-degenerate.
The two combinations of the signal photon in $S$ and idler in $A$ ($SA$) and the idler in $S$ and signal in $A$ ($AS$) are degenerate, as we consider a type-I PDC process where the generated photons are indistinguishable \cite{loudon_quantum_2000}.
This leads to the formation of three spectrally distinct phase-matching curves, as shown in Fig. \ref{fig:sys_sketch}(c). Due to this spectral separation, we can selectively excite different eigenmode combinations by choosing the correct pump wavelength.

Selecting one of the three phase-matching conditions imprints specific spatial properties on the generated photon pairs. This intrinsic feature of the system allows us to engineer different quantum states of light by tuning the pump wavelength regardless of the spatial distribution of the pump. However, although the spatial properties of the state do not depend on the length of the device $L$, we need a minimum length of the poled region to eliminate spectral overlap between different phase-matching conditions to get a clear spatial signature of the state.

We generate the postprocessing free NOON state in the waveguide basis by exciting only the central phase-matching condition, which corresponds to the generation of one photon in each eigenmode. This can be shown by rewriting Eq. \eqref{eq:eig_state} in the waveguide basis considering only the $\Delta\beta_{AS,SA}$ contributions
\begin{equation}
\begin{aligned}
\ket{\Psi}^\mathrm{wg}&=\frac{\delta}{2\sqrt{\mathcal{N}}}\int \mathrm{d}\omega_s \int\mathrm{d}\omega_i\, \alpha(\omega_s+\omega_i)\\
&\left[ \left\{\Phi(\Delta\beta,L)+ \Phi(\Delta\beta,L)\right\}\h{b}^\dagger_1(\omega_s) \h{b}^\dagger_1(\omega_i)\right. \\
&- \underbrace{\left\{ \Phi(\Delta\beta,L)- \Phi(\Delta\beta,L)\right\}}_{=\,0}\h{b}^\dagger_1(\omega_s) \h{b}^\dagger_2(\omega_i) \\
&+\underbrace{\left\{ \Phi(\Delta\beta,L)-\Phi(\Delta\beta,L)\right\}}_{=\,0}\h{b}^\dagger_2(\omega_s) \h{b}^\dagger_1(\omega_i) \\
&-\left\{\left.\Phi(\Delta\beta,L)+ \Phi(\Delta\beta,L)\right\}\h{b}^\dagger_2(\omega_s) \h{b}^\dagger_2(\omega_i) \right] \ket{0}\\
&=\frac{\delta}{\sqrt{\mathcal{N}}} \int \mathrm{d} \omega_s \int \mathrm{d} \omega_i \, \alpha(\omega_s+\omega_i)\, \Phi(\Delta\beta,L)\\
&\times \left[\h{b}_1^\dagger(\omega_s)\h{b}_1^\dagger(\omega_i)-\h{b}_2^\dagger(\omega_s)\h{b}_2^\dagger(\omega_i)\right]\ket{0}\, ,
\end{aligned}
\label{eq:wg_state}
\end{equation}
where we have substituted $\Delta\beta_{SA}=\Delta\beta_{AS}=\Delta\beta$. The structure of this state clearly shows that we generate a genuine postprocessing free two-photon NOON state in our device, i.e., both photons exit the chip in either waveguide in a coherent superposition. The key to the NOON state generation is embedded in the cross terms $\hat{b}_{1,2}$ of this state. We have already stated that it is possible to simultaneously generate one photon (signal) in the symmetric eigenmode and the other (idler) in the antisymmetric eigenmode. However, the interchanged combination (idler in the symmetric and signal in the antisymmetric) is also possible, but with a phase flip. As these two possibilities are indistinguishable, the two terms cancel out during the basis transformation, which results in the postprocessing free two-photon NOON state.

From a physics point of view, the main point of this scheme is the choice of the waveguide modes as a natural detection basis and the transformation from eigenmodes to waveguide modes as sketched in Fig. \ref{fig:sys_sketch}(e). The linear transformation between eigenmode and waveguide basis is mathematically fully equivalent to a perfect 50:50 beam splitter. In this analogy, the two input ports of the beam splitter represent the two eigenmodes and the two waveguide modes compose the output ports. As we generate one photon in each eigenmode, the combination of both photons in the basis transformation gives rise to Hong-Ou-Mandel-like interference \cite{hong87prl}, which extinguishes the probability of detecting one photon in both waveguides. Note, however, that there is no \textit{physical} beam splitter implemented on the chip. Only the choice of waveguide modes as a measurement basis and the corresponding basis transformation give rise to the beam splitter equivalent working on our quantum state.

\section{Chip Design}
For the implementation of the NOON state source, we realize the waveguide coupler structure in lithium niobate. The high nonlinearity of the material, as well as the reliable fabrication of waveguides by titanium indiffusion, results in high-quality quantum optical devices. 
The fabricated Ti:LN waveguides have a width of $6.9\,\mu$m and a center-to-center separation in the coupling region of $13\,\mu$m. The coupling region of $L=11\,$mm is enclosed with two bending regions that bridge a waveguide separation of $100\,\mu$m at the incoupling facet and a separation of $165\,\mu$m at the out-coupling part of the chip. We optimised the titanium layer thickness for the indiffusion process to $79\,$nm to minimize cross coupling in the bendings after the coupling region of the chip. After the waveguide fabrication we performed a periodic poling of the chip with a grating period of $\Lambda_{pp}=16.6\,\mu$m. These fabrication parameters lead to a coupling parameter of $C=(358\pm10)\,\mathrm{m}^{-1}$ and a phase-matching condition for $\Delta\beta_{SA}=\beta(759.7\,\mathrm{nm})-\beta(1519.4\,\mathrm{nm})-\beta(1519.4\,\mathrm{nm})=0$ at room temperature. Furthermore, the high quality of the waveguides manifests itself in the low loss values of $\alpha=0.2...0.5\,\mathrm{dB}\mathrm{cm}^{-1}$.

\section{Experimental Setup \& Results}
\begin{figure}
\includegraphics[width=1.\columnwidth]{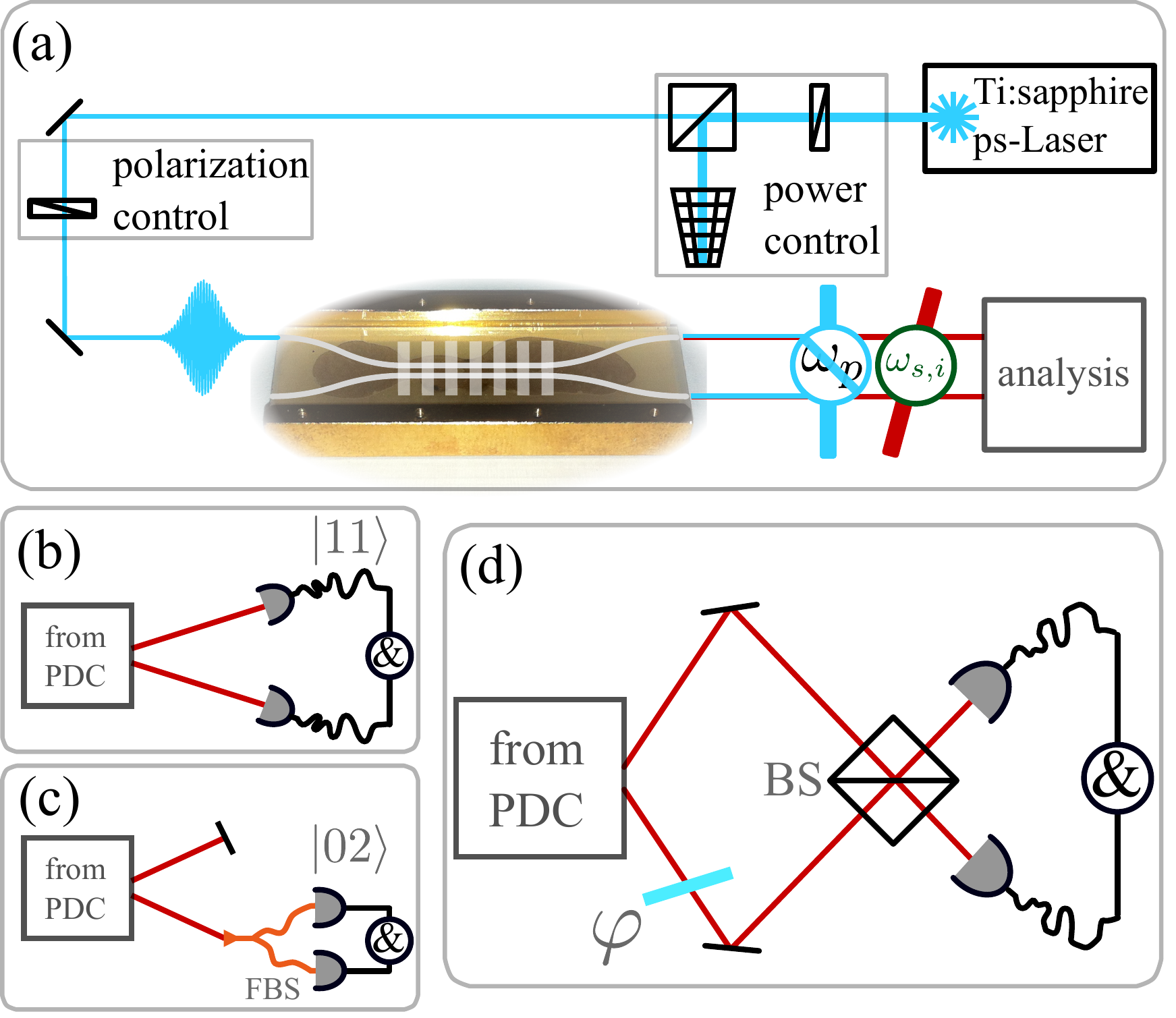}
\caption{(Color online) Setup. (a) We pump a type-I PDC source at room temperature with a picosecond-pulsed laser system and couple the beam only to a single waveguide of the device. After the chip, we filter out the pump light and the undesired background and send the generated photon pairs to the analysis part. (b) We measure the two-photon events between waveguides ($\ket{1,1}$) by coupling the waveguide outputs directly to detectors. (c) By inserting a fiber-based 50:50 beam splitter (FBS) we measure the coincidences in a single waveguide ($\ket{2,0}$ and $\ket{0,2}$). (d) To test the phase coherence between the two waveguide outputs, we interfere the two paths on a bulk beam splitter (BS) and change the relative phase $\varphi$ with a thin glass plate.
}
\label{fig:setup}
\end{figure}

To demonstrate the generation of indistinguishable two-photon NOON states in the fabricated device, we adopted the experimental setup as shown in Fig. \ref{fig:setup}. We pump our source with a picosecond-pulsed Ti:sapphire laser source with a repetition rate of $R_{\mathrm{rep}}=1\,$MHz, followed by a power and polarization control as depicted in Fig. \ref{fig:setup}(a). The pulsed operation of our source coupled with the low repetition rate allows us to avoid the photorefractive effect \cite{weis_lithium_1985} at room temperature due to a long relaxation time between two consecutive pulses. After the PDC takes place in the coupler structure, we send the generated photons together with the remaining pump light to a filtering stage, where the pump is suppressed and we remove unwanted background in the telecom regime with a broad band filter that has a bandwidth of $50\,$nm. We characterized the high brightness of our device and found an estimated time-averaged generation rate of $1.5\times 10^5\mathrm{pairs}\mathrm{s}^{-1}\mu\mathrm{W}^{-1}$, which proves the high quality of our fabrication technique. To analyze the two-photon detection events (coincidences) in the waveguide basis, we first probe the $\ket{11}$ contribution by coupling the waveguide modes directly to two avalanche photodiodes as depicted in Fig. \ref{fig:setup}(b). We measure the coincidences in a single waveguide ($\ket{20}$ and $\ket{02}$) by coupling one waveguide to a fiber and insert a 50:50 fiber-based beam splitter in Fig. \ref{fig:setup}(c).

\begin{figure}
\includegraphics[width=1.\columnwidth]{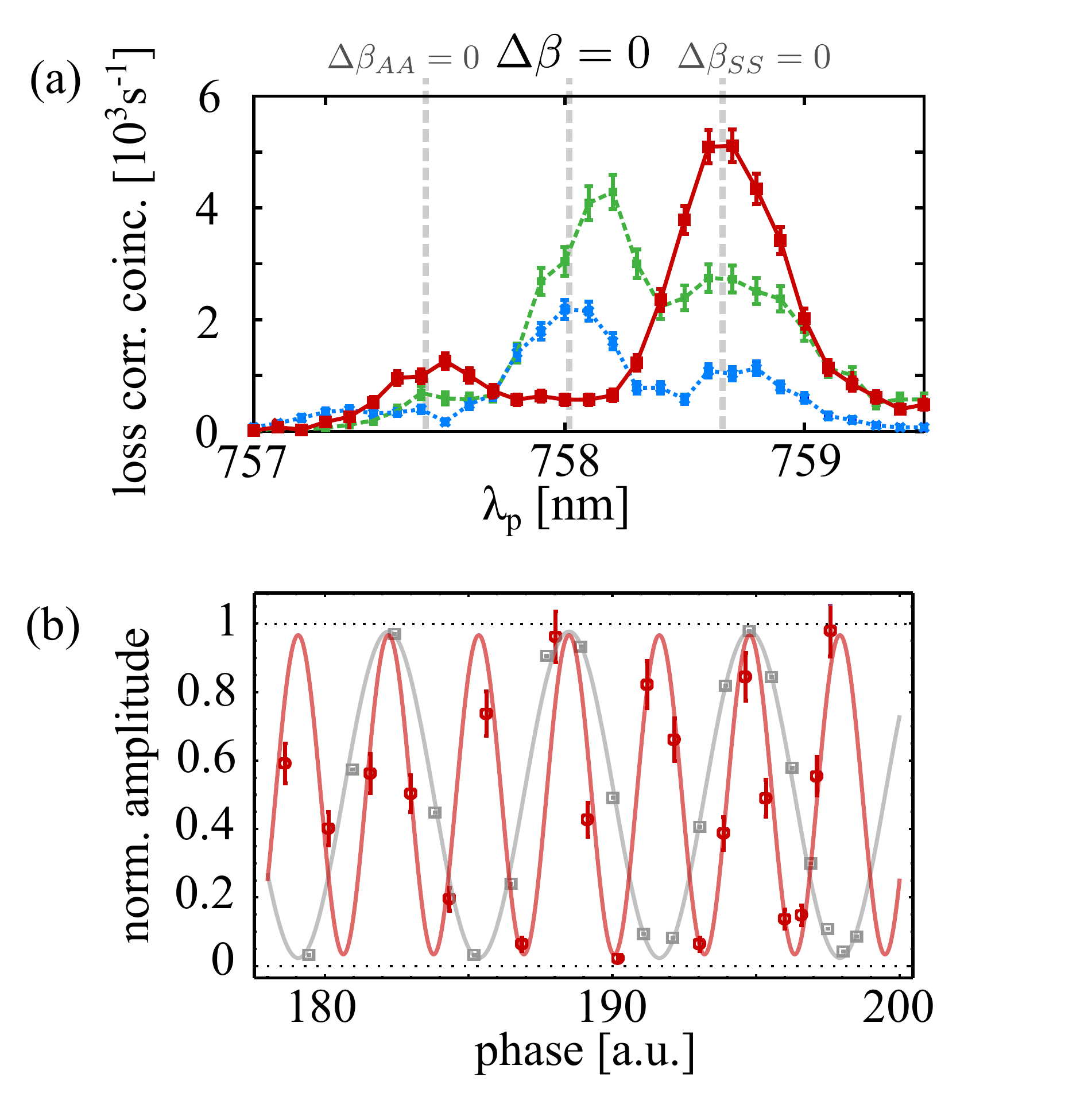}
\caption{(Color online) Results. (a) To find the working point of the source, we tune the pump wavelength and record the coincidence events (green, pumped waveguide; blue, unpumped waveguide; and red, between waveguides). We find a fidelity of $\mathcal{F}=(84.2\pm2.6)\,\%$ at $\lambda_p=758.1\,$nm. (b) For the coherence test, we vary the phase with a small glass plate and find the expected double fringe pattern of a two-photon NOON state (red) compared to a classical reference (gray). The high visibility of $\mathcal{V}_{\mathrm{N00N}}=(93.3\pm 3.7)\,\%$ shows that our photons are indistinguishable.
}
\label{fig:results}
\end{figure}

For a perfect NOON state, we expect coincidences in either waveguide with full suppression of coincidences between the waveguides. As we need a specific selection of the pump wavelength for the generation of the two-photon NOON state, we tune the pump wavelength and scan the complete phase-matching function to find the correct pump wavelength for the NOON state generation. The result of this measurement is plotted in Fig. \ref{fig:results}(a). By tuning the pump wavelength we find the signature of the three expected phase-matching conditions for different eigenmode combinations. The asymmetry between the different phase-matching conditions is an artifact of inhomogeneous periodic poling, however it does not affect the performance of this device. In the center of the figure at roughly 758$\,$nm, corresponding to a phase-matching condition of $\Delta\beta=0$, we find a suppression of coincidences between the waveguides (red curve), with enhancement of the event rates in the single waveguides (blue and green). This is the clear coincidence signature that is expected from a two-photon NOON state.

With this result, we calculate the expected fidelity of our state via
\begin{equation}
\mathcal{F}=\frac{R_1+R_2-R_{12}}{R_1+R_2+R_{12}}\, ,
\end{equation}
where $R_j$ are the coincidence counts in waveguide $j$ multiplied by 2 to correct for the 50:50 beam splitter and $R_{12}$ are the coincidences between waveguides 1 and 2. In our experiment, we find a measured fidelity of $\mathcal{F}=(84.2\pm2.6)\,\%$. For our implemented source the theoretically achievable maximum fidelity is $\mathcal{F}\approx93\,\%$, as it is influenced by the neighboring phase-matching conditions. However, this is not a fundamental restriction, as a longer coupler stem length or a narrower gap between waveguides reduces the side contributions significantly. The lower value obtained in the experiment is due to residual coupling between the two waveguide modes after the poled region, which enhances the effective $R_{12}$ contribution. However, it is possible to reduce this effect by careful fabrication of the waveguide structure after the poled region. 

To test the coherence properties of the state, we show the double-fringe pattern expected from a two-photon NOON state. We interfere the two waveguide outputs on a bulk 50:50 beam splitter as sketched in Fig. \ref{fig:setup}(d). In order to overlap the two waveguide outputs in time, we optimized the arm lengths of the interferometer by using a femtosecond-pulsed laser source  and achieved an interference visibility of $\mathcal{V}_{\mathrm{fs}}\approx95\,\%$, proving good overlap in both the spatial and time domains. For the classical reference of the fringing period, we use a continuous-wave laser at $\lambda_{\mathrm{coh}}=1520\,$nm, the same wavelength as the expected PDC light. We vary the phase between the two arms by tilting a thin glass plate and find an interference visibility of $\mathcal{V}_{\mathrm{coh}}=(95.4\pm0.4)\,\%$. The phase dependence of the reference is shown in Fig. \ref{fig:results}(b) (gray curve). The interference data with the PDC state are shown in red.
The doubled phase sensitivity \cite{dang01prl} is clearly observable in the PDC signal with respect to the classical reference. In the PDC measurement, we achieve a fringe visibility of $\mathcal{V}_{\mathrm{N00N}}=(93.3\pm 3.7)\,\%$ which fits the classically measured maximum visibility of the interferometer.

\section{Conclusion}
We have demonstrated a two-in-one waveguide source by exploiting the intrinsic generation protocol for nonlinear processes in coupled structures. We have shown how to harness the path degree of freedom in waveguide sources by using the eigenmodes for spatio-spectral engineering of PDC states in multiple channels. We found a Hong-Ou-Mandel effect in the transformation from generation to detection basis allowing for phase-stable state preparation by pumping only a single waveguide and independent of fabrication parameters.
Experimentally, we have also demonstrated reconfigurability of the quantum state in the spatial domain by only tuning the pump wavelength. Furthermore, we have shown the expected double-fringe pattern of a two-photon NOON state with a very high visibility, proving both the indistinguishability of the generated photons and the phase coherence between the two waveguide outputs.
Our approach eliminates additional overhead, such as narrow-band filtering or phase stabilization. 
While we have demonstrated the generation of two-photon NOON states in this work, expansion to higher-photon-number contributions shows that our device generates two fully identical, intrinsically-phase-stable squeezed states. This opens new perspectives in the field of integrated continuous-variable quantum optics. Moreover, combining the state generation with the action of linear elements takes the integration density to a new level, which is only possible due to the intrinsic $\chi^{(2)}$ nonlinearity of lithium niobate. This in turn simplifies the required complexity of linear quantum information networks, as the state preparation is already integrated in the source design and does not need any postprocessing by additional linear circuits.

\textit{Note added.} Recently, we became aware of parallel work by Setzpfandt \textit{et al.} \cite{setzpfandt_tunable_2015}.

\section*{Acknowledgments}
\vspace{-0.4cm}
R.K., L.S. and C.S. acknowledge funding by the DFG (Deutsche Forschungsgesellschaft) under Grant SFB/TRR 142 and from the European Union's Horizon 2020 research and innovation program under the QUCHIP project Grant No. 641039. C.S.H. and I.J. acknowledge support by the Ministry of Education of the Czech Republic Grant No. RVO 68407700 and the Czech Grant Agency (GACR) No. GACR 13-33906 S.\\

\clearpage

\end{document}